\documentclass[sigconf,screen]{acmart} 

\AtBeginDocument{%
  \providecommand\BibTeX{{%
    \normalfont B\kern-0.5em{\scshape i\kern-0.25em b}\kern-0.8em\TeX}}}

\let\oldtodo\todo
\usepackage{todonotes}

\let\todo\oldtodo

\usepackage{amssymb}

\usepackage[ruled,linesnumbered]{algorithm2e}
\usepackage{xcolor}
\usepackage{listings}
\usepackage{graphicx}
\usepackage{subcaption}
\usepackage{balance}
\usepackage{soul}
\usepackage{microtype}

\usepackage[final]{changes}

\lstdefinestyle{JavaStyle}{
    language=Java,
    basicstyle=\ttfamily\small,
    keywordstyle=\color{blue},
    commentstyle=\color{green!60!black},
    stringstyle=\color{orange},
    numbers=none,
    numberstyle=\tiny\color{gray},
    stepnumber=1,
    showstringspaces=false,
    breaklines=true,
    frame=single,
    backgroundcolor=\color{gray!5},
    captionpos=b,
    xleftmargin=2em,
}
\usepackage{multirow}
\usepackage{tcolorbox}
\usepackage{hyperref}
 \usepackage{enumitem}
\usepackage[]{collab}

\collabAuthor{yt}{blue}{Yintong Huo}
\collabAuthor{yx}{teal}{Yuxin Su}
\collabAuthor{yc}{gray}{Yichen}

\setcopyright{acmlicensed}
\acmDOI{10.1145/3650212.3652106}
\acmYear{2024}
\copyrightyear{2024}
\acmSubmissionID{issta24main-p42-p}
\acmISBN{979-8-4007-0612-7/24/09}
\acmConference[ISSTA '24]{Proceedings of the 33rd ACM SIGSOFT International Symposium on Software Testing and Analysis}{September 16--20, 2024}{Vienna, Austria}
\acmBooktitle{Proceedings of the 33rd ACM SIGSOFT International Symposium on Software Testing and Analysis (ISSTA '24), September 16--20, 2024, Vienna, Austria}
\received{16-DEC-2023}
\received[accepted]{2024-03-02}

\begin{document}


\title{Face It Yourselves: An LLM-Based Two-Stage Strategy to Localize Configuration Errors via Logs}

\author{Shiwen Shan}
\orcid{0009-0000-1317-8957}
\affiliation{%
  \institution{Sun Yat-sen University}
  \city{Zhuhai City}
  \country{China}
}
\email{shanshw@mail2.sysu.edu.cn}

\author{Yintong Huo}
\orcid{0009-0006-8798-5667}
\affiliation{%
  \institution{Chinese University of Hong Kong}
  \city{Hong Kong}
  \country{China}
}
\email{ythuo@cse.cuhk.edu.hk}

\author{Yuxin Su}
\authornote{corresponding author}
\orcid{0000-0002-3338-8561}
\affiliation{%
  \institution{Sun Yat-sen University}
  \city{Zhuhai City}
  \country{China}
}
\email{suyx35@mail.sysu.edu.cn}

\author{Yichen Li}
\orcid{0009-0009-8370-644X}
\affiliation{%
  \institution{Chinese University of Hong Kong}
  \city{Hong Kong}
  \country{China}
}
\email{ycli21@cse.cuhk.edu.hk}

\author{Dan Li}
\orcid{0000-0002-3787-1673}
\affiliation{%
  \institution{Sun Yat-sen University}
  \city{Zhuhai City}
  \country{China}
}
\email{lidan263@mail.sysu.edu.cn}

\author{Zibin Zheng}
\orcid{0000-0002-7878-4330}
\affiliation{%
  \institution{Sun Yat-sen University}
  \city{Zhuhai City}
  \country{China}
}
\email{zhzibin@mail.sysu.edu.cn}


\begin{abstract} 
    Configurable software systems are prone to configuration errors, resulting in significant losses to companies. However, diagnosing these errors is challenging due to the vast and complex configuration space. These errors pose significant challenges for both experienced maintainers and new end-users, particularly those without access to the source code of the software systems. Given that logs are easily accessible to most end-users, we conduct a preliminary study to outline the challenges and opportunities of utilizing logs in localizing configuration errors. Based on the insights gained from the preliminary study, we propose an LLM-based two-stage strategy for end-users to localize the root-cause configuration properties based on logs. We further implement a tool, LogConfigLocalizer, aligned with the design of the aforementioned strategy, hoping to assist end-users in coping with configuration errors through log analysis.
    
    To the best of our knowledge, this is the first work to localize the root-cause configuration properties for end-users based on Large Language Models~(LLMs) and logs. We evaluate the proposed strategy on Hadoop by LogConfigLocalizer and prove its efficiency with an average accuracy as high as $99.91\%$. Additionally, we also demonstrate the effectiveness and necessity of different phases of the methodology by comparing it with two other variants and a baseline tool. Moreover, we validate the proposed methodology through a practical case study to demonstrate its effectiveness and feasibility.
\vspace{-5pt}
\end{abstract}





\maketitle
\vspace{-5pt}
\section{Introduction} %
Configuration errors, also known as misconfiguration\added{s}, \replaced{are common and notorious anomalies in}{are a common and notorious anomaly for} configurable software systems. The term \replaced{refers to}{describes} the unexpected behavior \replaced{resulting from}{caused by} mistakenly setting an inappropriate value for a configuration property~\cite{yin2011empirical}
\replaced{, which poses a significant risk to software reliability}{. Mistakes in configuration pose a significant risk to software reliability}. The Open Web Application Security Project (OWASP)~\cite{owasp}, a community committed to trustworthy applications, identified configuration errors as a major vulnerability, ranking \added{the} fifth among the top ten in both 2021~\cite{owasp:2021} and 2022~\cite{owasp:2022} \deleted{editions?}.
\replaced{Configuration}{These} errors can significantly disrupt user experiences; for example, Sweden faced domain paralysis (\texttt{.se}) due to DNS configuration errors~\cite{example:se-sweden}, causing widespread inconvenience. Moreover, high-profile companies like Facebook, Microsoft Azure, and Amazon EC2 have \replaced{reported setbacks due to such errors~\cite{xu2013not}, indicating a widespread occurrence of configuration errors in the high-tech industries}{fallen victim to such errors~\cite{xu2013not}, highlighting the prevalence of configuration errors in the high-tech industry}. 

\replaced{Configuration errors are common software system anomalies, which are troublesome and particularly difficult to diagnose, even for experienced maintenance engineers, leading to significant side effects for companies, maintainers, and end-users~\cite{xu2015hey,zhang2013automated,xu2013not}.}{present significant challenges for companies and maintainers and end-users~\cite{xu2015hey,zhang2013automated,xu2013not}. 
These anomalies are widespread, troublesome, and particularly difficult to diagnose, even for experienced maintenance engineers.}
For end-users unfamiliar with configurable software, \replaced{comprehending}{understanding} and addressing these issues \replaced{could be even more}{can be further} daunting. 
However, existing \replaced{strategies, tackling configuration errors via program analysis, are predominantly designed for software developers rather than end-users who do not have access to source code ~\cite{zhang2014configuration,xu2016early,zhang2013automated,attariyan2010automating}.}{works primarily target program analysis aimed at developers rather than end-users~\cite{zhang2014configuration,xu2016early,zhang2013automated,attariyan2010automating}. These works often require access to software code, which might be inaccessible to end-users.} \replaced{On the other hand, even in cases with full access to source code}{Even when accessible}, pinpointing and resolving the root-cause configuration settings \deleted{through traditional program analysis} \replaced{remains}{is} a challenging and time-consuming task for end-users\replaced{. This challenge is posed by the extensive and intricate configuration space, exacerbated when there are dependencies or conflicts among the configuration properties}{due to the large and complicated configuration space~\cite{xu2015hey,yin2011empirical}, especially when there exists dependence or conflict relationship among the configuration properties}~\cite{xu2015hey,yin2011empirical,zhang2023fuzzing}.

Logs \replaced{with}{contain} software runtime information \added{are easily accessible for both software developers and end-users}, serving as a valuable resource for various software monitoring and failure diagnosis applications~\cite{he2021survey,cinque2012event,he2018identifying}. 
However, previous research has primarily focused on utilizing logs for system analysis~\cite{zhang2015proactive, 10.1145/3460319.3464799}, neglecting the potential of logs in pinpointing configuration errors. 
In this paper, we aim to bridge this gap by exploring how logs can be leveraged to automatically pinpoint configuration issues. To achieve this, we conduct a preliminary investigation \replaced{into}{on} the relation\added{ship} between the configuration errors and logs. \replaced{Through analyzing the 100 entries of posted configuration setting-related problems of Hadoop~\cite{hadoop} collected on Jira~\cite{jira} and StackOverflow~\cite{stack}, we identify opportunities to pinpoint the root-cause configuration properties by examining two types of anomaly symptoms present in the logs}{We collect 100 entries of posted configuration setting-related problems of Hadoop on Jira~\cite{jira} and StackOverflow~\cite{stack} We find the opportunities to localize the root-cause configuration properties by analyzing two types of symptoms in the logs}.   

In the preliminary investigation, we identify \deleted{that there are} two types of log symptoms \replaced{that indicate}{indicating} configuration errors.  \added{The direct symptom} directly presents the name or \deleted{the} value of the root-cause configuration property, \replaced{but matching such properties with logs requires fine-grained matching algorithms.}{while it calls for sophisticated algorithms to match the properties and logs;} \replaced{The indirect symptom involves a lack of direct information about the root-cause configuration property but pointing to other states of the system due to the invisible logic within the code.}{the other is in lack of specific information of the root-cause configuration property due to the invisible logic inside the code, but demonstrates other states of the system.} While both of the \replaced{aforementioned symptoms cannot be directly utilized to localize root-cause configuration properties}{log symptoms bring challenges to localize the root-cause configuration property}, they also bring opportunities. For the direct symptom, once the critical information is captured, we can directly localize the root-cause configuration property. For the indirect symptom, we can \deleted{still} \replaced{infer}{inference} the suspected root-cause configuration property by \replaced{comprehending}{understanding} and interpreting the related log messages.

Based on the insights from the preliminary study, we introduce an LLM-based two-stage strategy to localize the root-cause configuration property via logs.
The proposed methodology involves two stages, the \textit{Anomaly Identification Stage} and the \textit{Anomaly Inference Stage}. Given a set of logs and user-defined configuration settings, we first identify and select the log messages indicating configuration-related errors in the \textit{Anomaly Identification Stage}. Then we localize the suspected root-cause configuration properties based on the selected log messages and the offered configuration settings by introducing the rule-based phase and the LLM-based phases in the \textit{Anomaly Inference Stage}. To the 
best of our knowledge, it is the first work to locate configuration errors based on LLMs and logs.

\replaced{We demonstrate the performance of the proposed methodology by implementing a tool -- LogConfigLocalizer.}{We implement a tool, LogConfigLocalizer, to show the performance of the proposed methodology.} In addition, we establish a log benchmark containing various configuration errors by dynamically running five types of workloads with different configuration settings on Hadoop~\cite{hadoop}.
We show the high effectiveness of LogConfigLocalizer on the established benchmark\replaced{, achieving an average accuracy of 99.91\%}{with a mean accuracy of 99.91\%}. \replaced{Furthermore}{Additionally}, we compare LogConfigLocalizer with two variants and a baseline tool, \added{and} all \deleted{of the} experiments demonstrate the superior performance of LogConfigLocalizer. We further conduct a practical case study \replaced{to localize}{by localizing} the root-cause configuration properties of 33 cases \added{involved} in the preliminary study and demonstrate LogConfigLocalizer's feasibility with a high accuracy of 93.94\% (31/33). 

To conclude, our main contributions are listed as follows:

$\blacklozenge$ We conduct a preliminary study to explore the challenges and opportunities associated with localizing configuration errors through log analysis.

$\blacklozenge$ We introduce a two-stage strategy based on LLMs for end-users who are new to the software systems to localize the configuration errors.

$\blacklozenge$ We implement a tool, LogConfigLocalizer, to \replaced{assist}{give assistance to} end-users \replaced{in}{for} localizing configuration errors. The source code is publicly available\footnote{\url{https://github.com/shanshw/LogConfigLocalizer/}} to benefit future research. 

$\blacklozenge$ We demonstrate the effectiveness of LogConfigLocalizer with an average accuracy as high as $99.91\%$ in evaluations and show its feasibility through a practical case study. 

\section{Background}\label{sec:back}
\subsection{Problem Definition} 
In this paper, we formulate the log-based configuration error localization task as follows. Given a set of logs $L$ with $n$ log messages $L = \{l_1,l_2,...,l_n\}$, and user-defined configuration settings $C_u$ with $m$ entries $C_u = \{e_1,e_2,...,e_m\}$, where an entry $e_i=(p_i,v_i)$ indicates a specific configuration property $p_i$ and value $v_i$, the output is formatted as a key-value set $S$ with $t$ entries containing the suspected configuration error triggers $e_{sj}=(p_{sj},v_{sj})$, $S = \{e_{s1}, e_{s2}, ..., e_{st}\}$.

Moreover, we clarify the frequently used terms in the paper.
\begin{itemize}[leftmargin=*]
    \item \textit{May-Fault Logs}: Logs offered by end-users that may contain configuration errors.
    \item \textit{Fault-Free Logs}: Logs generated with conventional configuration settings, namely the minimum configuration settings for an application to run. Fault-Free is short for Configuration-Fault-Free.
    \item \textit{Configuration Error Triggers}: The most likely configuration settings to trigger a configuration error. 
\end{itemize}
Notably, the core idea of our methodology is similar to the signature-based approaches, which attempt to localize the configuration errors by comparison between the offered signature and the reference signature~\cite{xu2015systems}. A signature refers to the runtime information of systems, such as system call traces~\cite{yuan2006automated}. The offered signature is generally considered to record the target failure information and the reference signature can record either normal or known failure information~\cite{xu2015systems}. In this case, the input log files containing configuration errors (i.e., may-fault logs) can be regarded as the offered signature and fault-free logs can be seen as the reference signature.

\begin{table*}[htb!]
\small
  \caption{Statistics of the Preliminary Study. \#w/ Log shows the number of problems posted with logs, \#S-* indicates the two symptoms in reported logs, and \#T-* denotes the configuration error type. Some posted problems cover all the configuration properties, thus not included in the configuration error type measurement. }
  \vspace{-5pt}
  \label{tab:preliminary_statistic}
    \begin{tabular}{c|c|c|c|c|c|c|c|c|c} 
    \toprule
    {} &  \textbf{\#w/ Log} & \textbf{\#S-Direct}  & \textbf{\#S-Indirect} & \textbf{\#T-Path} & \textbf{\#T-Numeric} & \textbf{\#T-Classpath} & \textbf{\#T-Boolean} & \textbf{\#T-String} &\textbf{\# Total}  \\ 
    \hline
    \textbf{Jira} &  41  & 13 & 28 & 10 & 30 & 4 & 9 & 13 & 68 \\
    \hline
    \textbf{StackOverflow} & 23 & 0  & 23 & 9 & 4 & 6 &2 & 5 & 32\\
    \hline
    \textbf{Percentage} & / & / & / & 20.65\% & 36.96\% & 10.87\% & 11.96\% & 19.57\% & 100\% \\
    \bottomrule
  \end{tabular}\\
\end{table*}
\subsection{Preliminary Study} %
We conduct a preliminary study to consider the possibilities of using logs to localize configuration errors by investigating the relationship between configuration errors and logs.

To begin with, we select the distributed big data framework, Hadoop~\cite{hadoop}, as our study system, collecting 100 configuration error reports submitted to Jira~\cite{jira} or StackOverflow~\cite{stack}, which contain numerous technical discussions on software system runtime errors.

For selection, we initially search for reports using keywords (e.g., "configuration", "error", "failure" and the names of configuration properties). We then select the top 100 cases returned by the websites, upon confirming they are related to configuration errors. Such a decision is made by reviewing their comments and descriptions. If the error is resolved by adjusting the configuration settings, we consider this report highly relevant. 

The reports record detailed failure information, including configuration error triggers, the inappropriately-set values, the data type of the values, and logs. Table~\ref{tab:preliminary_statistic} shows the statistics of these reports.


\subsubsection*{\textbf{A. How many configuration error reports contain logs?}}\label{sec:prea} 
We manually count the number of reports containing logs to examine whether they play an important role in configuration error diagnostics. According to Table~\ref{tab:preliminary_statistic}, more than half (64\%) of end-users share application logs to detail configuration errors. 
Additionally, some (5/100) developers in StackOverflow~\cite{stack} further request logs from users to pinpoint anomalies if the initial logs are insufficient.
Notably, some end-users describe the system's unexpected behavior in natural language, leading to reports lacking logs.
\begin{tcolorbox}[
    colback=orange!5!white,
    colframe=orange!75!black,
    title=Finding 1,
    fonttitle=\bfseries,
    sharp corners
]
The majority (64\%) of end-users attach logs in their anomaly reports, revealing the value of logs for diagnosing configuration errors.
\label{box:fd1}
\end{tcolorbox}

\subsubsection*{\textbf{B. How do logs reflect configuration errors?}} 
We conduct a detailed analysis of \textit{anomaly symptoms} inside end-user logs and classify them into two categories: \textit{direct} and \textit{indirect}.
The direct symptom specifies the name or value of the root-cause configuration property. 
In contrast, indirect symptoms lack explicit information about the root-cause configuration property, but instead reveal additional system run-time behaviors, such as the stack statements.
Among 64 reports with logs examined, 20\% (13/64) logs exhibit the direct symptoms and the others show the indirect symptoms. However, we believe that the direct symptoms shall occur more often in practice, given that it is convenient for end-users to inspect log records and use the identified information (e.g., the name or the value of the root cause configuration property) to rectify their errors directly. 
\begin{tcolorbox}[
    colback=orange!5!white,
    colframe=orange!75!black,
    title=Finding 2,
    fonttitle=\bfseries,
    sharp corners
]
Logs reveal anomaly symptoms in two ways: direct and indirect. The direct symptom indicates the root-cause configuration properties while the indirect one lacks explicit information. They occupy 20\% and 80\% of cases in our study, respectively.
\label{box:fd2}
\end{tcolorbox}
\begin{figure}[htb!]
    \centering
    \includegraphics[width=0.48\textwidth]{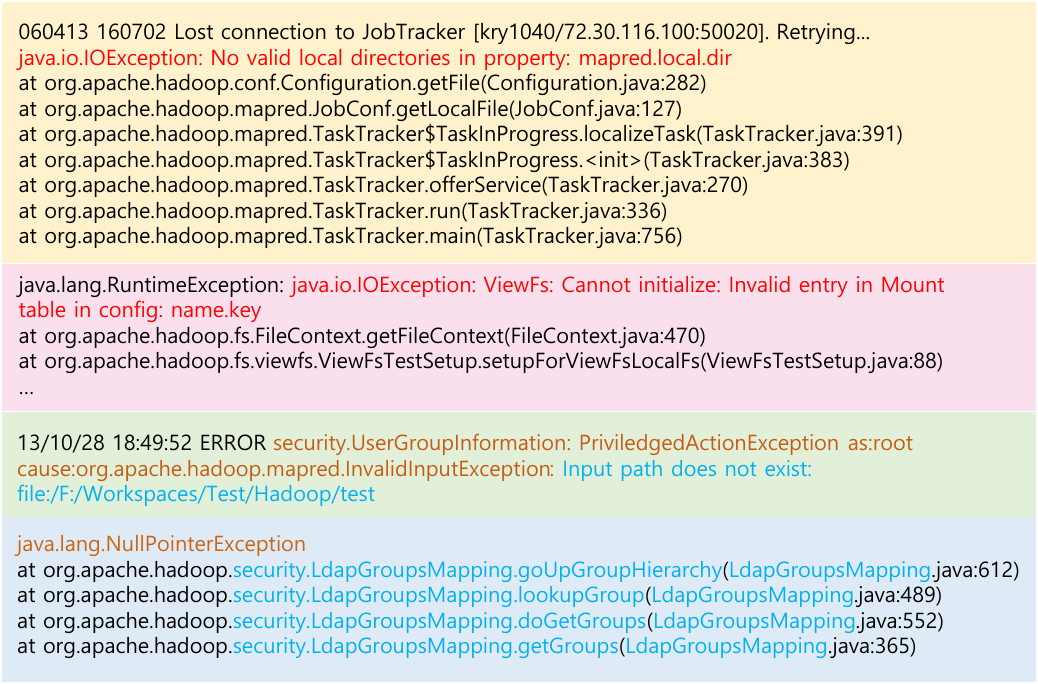}
    \vspace{-20pt}
    \caption{Two Types of Anomaly Symptoms in Logs} 
    \label{fig:two_tyoe_log_sym}
\end{figure}
Figure~\ref{fig:two_tyoe_log_sym} shows examples of the two symptoms. Specifically, the upper two boxes are examples of the direct symptoms, and the lower two indicate the indirect symptoms.

The direct symptoms present the explicit information of the configuration error triggers. The content enclosed in the orange box\footnote{Original report: \url{https://issues.apache.org/jira/browse/HADOOP-134}.} represents the direct symptom showing the full name of the configuration error trigger highlighted in red log  (i.e., \texttt{mapred.local.dir}). 
In this case, end-users can localize the configuration error triggers by directly matching the presented name or value. 
However, some configuration error triggers in direct symptoms lack their fully qualified names, leading to a matching challenge. This deficiency in information potentially stems from the complexity of long property names.
The pink box~\footnote{Original report: \url{https://issues.apache.org/jira/browse/HADOOP-18802}.} exemplifies this challenge.
Instead of displaying the full name of the property \texttt{fs.viewfs.mounttable.default.\\name.key}, only partial fragments like \texttt{name.key} are presented in the red log.
One potential approach to tackling this challenge involves devising fuzzy matching strategies. By breaking down the full name into sub-names (\texttt{name}, \texttt{key}) and utilizing these components to match corresponding keywords in the logs, there remains a possibility of identifying the configuration error trigger.
\begin{tcolorbox}[
    colback=orange!5!white,
    colframe=orange!75!black,
    title=Finding 3,
    fonttitle=\bfseries,
    sharp corners
]
The direct symptoms pose an insufficient information challenge, which requires a fine-grained matching algorithm to address it.
\label{box:fd3}
\end{tcolorbox}
Indirect symptoms provide no specific details about configuration error triggers but offer alternative insights into the runtime behavior of software systems.
The orange text within the green box\footnote{Original report: \url{https://stackoverflow.com/questions/19636220/exception-while-submitting-a-mapreduce-job-from-remote-system}.} indicates job submission failures without explicit configuration error triggers. However, the blue-highlighted text indicates the anomaly's manifestation -- a nonexistent path. 
This observation suggests a potential connection between the configuration error triggers and directories/path settings. 
Notably, the presented path is not part of the user-defined configurations and the absence of a value for \texttt{mapred.local.dir} introduces this anomaly.

The case within the blue box~\footnote{Original report: \url{https://issues.apache.org/jira/browse/HADOOP-18821}.} displays the value of stack statements in logs.
Following the \texttt{NullPointerException}, these stack statements provide clues about the root-cause configuration property\footnote{\texttt{hadoop.security.group.mapping.ldap.search.group.hierarchy.levels}}. By tracing a sequence within LdapGroups Mapping, these statements suggest that the security settings related to LdapGroups Mapping might be the root cause.


The indirect symptoms manifest the anomaly without explicitly identifying the culprits. It presents a challenge for end-users to intuitively correlate these anomalies with their configuration settings via logs. Such a challenge arises due to the discrepancy between the natural language used in logs and the structured configuration settings.
However, we can still infer configuration error triggers by comprehending and interpreting the alternative information in the indirect symptoms. 
Besides, the stack statements recording method invocation sequences, offer an alternative view to identify configuration error triggers.
\vspace{-2pt}
\begin{tcolorbox}[
    colback=orange!5!white,
    colframe=orange!75!black,
    title=Finding 4,
    fonttitle=\bfseries,
    sharp corners
]
The discrepancy between the natural language used in logs and structured configuration settings presents challenges with indirect symptoms, requiring deeper comprehension of alternative information.
\label{box:fd4}
\end{tcolorbox} 

\subsubsection*{\textbf{C. What's the most error-prone data types?}}
New users of software systems often make configuration mistakes due to complicated configuration properties with unclear or absent descriptions~\cite{xu2015hey, yin2011empirical}. To better understand what configuration properties confuse end-users the most, we further explore the types of incorrectly configured values. The statistical results, presented in Table~\ref{tab:preliminary_statistic} with columns prefixed by "\#T-", review five distinct types of values in the examined cases: "Path", "Numeric", "Classpath", "Boolean” and “String”. 
"Path" includes paths to existing directories and IP addresses with specified ports; "Numeric" represents numerical values; "Boolean" indicates true or false states; "String" denotes the system-recognizable names of components; and "Classpath" signifies the fully qualified name of a Java class. 
\begin{tcolorbox}[
    colback=orange!5!white,
    colframe=orange!75!black,
    title=Finding 5,
    fonttitle=\bfseries,
    sharp corners
]
Concerning the misconfigured data types, numeric values notably trigger common configuration errors by making up 37\%, while path and string types each represent 20\%. Boolean accounts for 12\%, and Classpath comprises 11\%.
\label{box:fd5}
\end{tcolorbox}

\begin{figure*}[htb!]
    \centering
    \includegraphics[width=\textwidth]{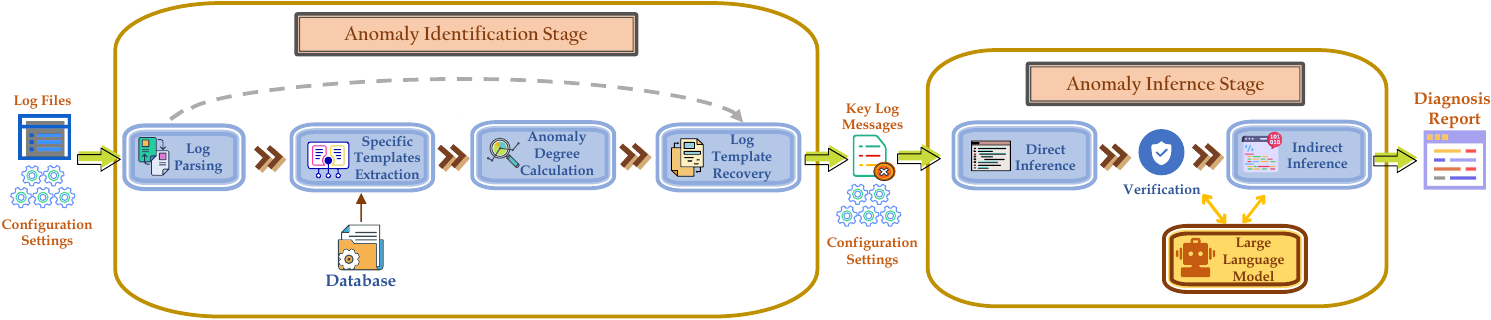}
    \caption{Overview of the LLM-based Two-Stage Strategy} 
    \label{fig:overview}
\end{figure*}
\vspace{-10pt}
\section{Configuration Bug Localization}\label{sec:method} 
\subsection{Overview}

Inspired by the value of logs from the preliminary study, we introduce an LLM-based two-stage strategy. Figure~\ref{fig:overview} illustrates the framework of the two-stage strategy for configuration-related error localization through log analysis. 

The first stage is the \textit{Anomaly Identification Stage}. The stage aims to identify the log messages related to configuration errors.
To begin with, we parse end-users' logs into log templates. These templates become the basis to identify the specific templates in may-fault logs, by comparing them with fault-free logs stored in our database. 
We then compute the anomaly degree for each specific log in may-fault logs. The logs, whose templates receive an anomaly degree greater than zero, will be identified as ``key log messages'' and progress to the next stage. Otherwise, these may-fault logs are identified as configuration-fault-free, concluding the localization process. 

For the \textit{Anomaly Inference stage}, the key log messages will proceed to the \textit{Direct Inference Phase} initially, attempting to directly localize the configuration error triggers based on rules. If successful, the inferred configuration error triggers will be passed to the \textit{LLM-powered Verification Phase} for verification. However, failures in either the \textit{Direct Inference Phase} or the \textit{Verification phase} will redirect the flow of the second stage towards the \textit{LLM-based Indirect Inference Phase}. A diagnosis report will be generated for end-users at the end of the localization procedure.   
The following sections provide illustrations for each stage.

\subsection{Anomaly Identification Stage} 
The extensive logs hold information revealing configuration errors, but they also include additional runtime information, involving the software resource utilization and the states of running jobs. Therefore, we devise a rule-based strategy with four phases in the stage to identify anomalous may-fault logs and distinguish configuration error-related logs from other operation logs.

\subsubsection{Log Parsing} 

Run-time logs are semi-structured data, consisting of constant strings and run-time variables. Parsing these logs involves extracting constant strings, known as log templates, and replacing run-time variables with placeholder symbols~\cite{zhang2019robust,5360240,9251078,du2017deeplog,huo2023semparser}. Leveraging log templates as representative patterns can alleviate the subsequent analysis workload and reduce associated costs~\cite{du2017deeplog,he2018identifying,he2021survey, huo2023autolog}.

In the early phase, we adopt log parsing to turn the may-fault logs into log templates. However, run-time variables provide valuable resources, allowing us to explore software system status further~\cite{9678773,huo2022logvm}. Therefore, we preserve the filtered run-time variables for the following recovery phase. To illustrate, in Figure~\ref{fig:overview}, the gray dashed arrow from the \textit{Log Parsing Phase} to the \textit{Log Template Recovery Phase} represents the flow of the stored run-time variables.

\subsubsection{Specific Templates Extraction}
Assuming that fault-free logs do not contain configuration errors, we leverage the distinct hash codes of fault-free log templates and extract them from may-fault log templates. 
Any template whose hash code does not match those in our database is considered specific, ensuring both speed and precision in identifying unique log templates.

\subsubsection{Anomaly Degree Calculation}
Merely identifying logs with specific templates as anomalies can be overly simplistic and may lead to numerous false positives. 
To mitigate the problem, we incorporate a heuristic anomaly degree calculation algorithm to further discern the anomalous log templates. 
Algorithm~\ref{first_algorithm} demonstrates the anomaly degree calculation procedure. To calculate the anomaly degree on a given log template, we have a weighted token set $S$, containing key tokens $t$ revealing anomalous and erroneous information. Each token weights differently during the calculation. Both the selection of tokens and their assigned weights are customizable, accommodating the diverse characteristics of various software systems. 
\vspace{-10pt}
\begin{algorithm}
  \caption{Anomaly Degree Calculation Algorithm}
  \label{first_algorithm}
  \KwData{Log Template $L$, Weighted Token Set $S$}
  \KwResult{Anomaly Degree $D$ of $L$}
  Initialize $D$ to zero\;
  \For{each token $t$ in $S$}{
    \If{$t$ exists in $L$}{
      $D$ += weight of $t$\;
    }
  }
  \Return $D$\;
\end{algorithm}

\vspace{-15pt}
In this phase, we designate may-fault logs as anomalous based on the anomaly degree of each log template. A higher anomaly degree (greater than zero) associated with a log template implies the presence of at least one log message offering additional anomaly details. Therefore, if any specific log template exhibits an anomaly degree beyond zero, the corresponding may-fault logs are identified as anomalous. As this phase concludes, for may-fault logs identified as configuration-error-free, the localization procedure ends with a diagnosis report indicating no configuration error occurs. Conversely, logs marked as anomalous progress to the subsequent phase for further examination.

\vspace{-5pt}
\subsubsection{Log Template Recovery}
A specific log template may correspond to numerous log messages, which could potentially increase the complexity of the subsequent stage. 
Therefore, we recover the filtered log templates to the log messages with the highest anomaly degree, referred to as "key log messages".

As previously emphasized, run-time variables are significant. 
Hence, to calculate the highest anomaly degree of a log message, we take the anomaly degree of the run-time variables into account. 
To be detailed, for a given filtered log template, we retain the corresponding log message whose run-time variables score the highest anomaly degree. 
These recovered log messages (i.e., key log messages) will be passed to the \textit{Anomaly Inference Stage} for configuration error localization.


\vspace{-10pt}
\subsection{Anomaly Inference Stage}
We propose a tri-phase strategy to localize the configuration error triggers with the set of key log messages generated in the \textit{Anomaly Identification Stage}. 

\subsubsection{Direct Inference} 
From Finding 2 and Finding 3, we recognize an opportunity for a direct search within logs for property names or values. 
This leads us to propose the \textit{Direct Inference Phase}.  

Given a set of key log messages, the \textit{Direct Inference Phase} is proposed to pinpoint configuration error triggers by directly matching property names or values. 
Intuitively, we leverage two methods: one focusing on property names and the other on property values.
When dealing with property name matching, a challenge lies in directly matching the entire name. To address this, we segment property names into distinct items based on the period (i.e., .), utilizing these segments for an exact match algorithm across logs.
Moreover, we exclude certain commonly used terms specifying components, such as "hadoop" for Hadoop~\cite{hadoop}, to reduce false positives. 
To achieve this, we extract configuration properties from the official-provided user manuals and documentation and break them into individual items. Subsequently, we count the occurrences of each item and form the filter set consisting of the top 20 items. 
Regarding property value matching, we directly search the value in the logs, following practices in previous works~\cite{zhang2015proactive,10.1145/3460319.3464799}.

The \textit{Direct Inference Phase} is sound when logs exhibit indicators, namely the information regarding the full or partial property name and the corresponding value. 
However, false positives may still occur despite the exclusion of hot terms. For instance, in the value matching strategy, a numeric value within a log message can represent various entities such as an IP address, a port, retry attempts, and so forth. 
The matched values do not always stand for the property value. To alleviate it, the following \textit{LLM-powered Verification Phase} is introduced.

\subsubsection{LLM-powered Verification} 


We present the \textit{LLM-powered Verification Phase} to address false positives introduced during the \textit{Direct Inference Phase} and ensure a second chance in case of a slip-up in the \textit{Direct Inference Phase}. It guarantees the overall accuracy of the \textit{Anomaly Inference Stage}. 

Logs, presented in natural language, are crafted for human readability. However, comprehending logs can be challenging for end-users and even experts in some cases. Therefore, we utilize Large Language Models (LLMs) for verification, aiming to harness the robust comprehension capabilities inherent in LLMs. 
We formulate a binary classification task for LLMs. Specifically, we furnish it with paired entries from the \textit{Direct Inference Phase}, such as the matched log message along with the corresponding matched configuration property and its value. In addition, for each matched property, we simultaneously provide its descriptions to LLMs for better performance. The description of each property is accessible in the documentation and the user manual. Next, we outline the binary classification task for the LLM, instructing it to generate output in a predefined format. If there is at least one configuration error trigger deemed plausible, we consider the results from \textit{Direct Inference Phase} to pass the \textit{Verification Phase}, concluding the workflow by generating a diagnostic report for end-users. Otherwise, it proceeds to the \textit{LLM-based Indirect Inference Phase} for a second attempt.

\vspace{-5pt}
\subsubsection{LLM-based Indirect Inference}\label{sec:two_cases}
The capability of the \textit{Direct Inference Phase} is limited to the direct symptoms, while it requires expertise and log-understanding ability to utilize the indirect symptoms as Finding 4 shows. 
Recently, LLMs have shown their strong power in natural language understanding and processing, hence we introduce the \textit{LLM-based Indirect Inference Phase} to grasp a second opportunity to localize the configuration error triggers.

Specifically, the localization procedure enters the \textit{Indirect Inference Phase} in two cases. The first case occurs when the \textit{Direct Inference Phase} fails, resulting in no matched entry being generated, prompting the workflow to skip the \textit{Verification Phase} and proceed directly to the \textit{Indirect Inference Phase}. It is referred to as a \textit{direct flow}. 
The second case arises from a failure in the \textit{Verification Phase}, indicating the localization procedure will go through all the phases in this stage, hence termed a \textit{complete flow}. 
In the \textit{direct flow}, no prior judgment is made, while in the \textit{complete flow}, failure in the \textit{Verification Phase} suggests a lack of trustworthiness in the \textit{Direct Inference Phase}. 
Therefore, it is reasonable to leave out information from the previous phases and utilize the logs, configuration settings, and descriptions instead.
We delegate the task of identifying configuration error triggers to LLMs. 
We provide them with details on key log messages, complete configuration settings, and related descriptions. 
To ensure effectiveness, We limit the number of the suspected configuration error triggers inferred by the LLMs.
In addition, we request LLMs to provide explanations for each selected suspected configuration error trigger.
We design the strategy to increase the reliability of LLMs' judgments and to provide end-users with additional information about the configuration errors.
\vspace{-8pt}
\section{Implementation}\label{sec:imple}
We implement LogConfigLocalizer based on the design of the proposed methodology. The following sections show the details.
\vspace{-8pt}
\subsection{Log Parsing}
We use Drain~\cite{8029742}, a prominent log parsing algorithm, which uses a fixed depth parse tree to expedite the parsing process for log parsing. It typically skips unrelated log file lines, including stack statements, which are crucial in the \textit{Anomaly Inference Stage}. Thus, we introduce an enhanced version of Drain~\cite{8029742} that includes stack statements. In the \textit{Direct Inference Phase}, we exclude stack statements to minimize false positives, reserving them for use in the \textit{Indirect Inference Phase} to provide more detailed information to LLMs.

\vspace{-10pt}
\subsection{Anomaly Degree Calculation}
It's intuitive to consider diagnostic information when selecting tokens. 
Therefore, we establish the set $S$ with tokens: \textit{error}, \textit{exception}, \textit{invalid}, \textit{failure}, \textit{disable}, \textit{false}, \textit{fault}, \textit{warn}, \textit{because} and \textit{exit}. 
By default, each token is assigned an equal weight of 0.1, adhering to an equal allocation strategy. This approach ensures the process remains domain-agnostic, requiring no specific domain knowledge for its implementation.
\vspace{-5pt}
\subsection{LLM Selection}
\begin{figure}
    \centering
    \includegraphics[width=0.4\textwidth]{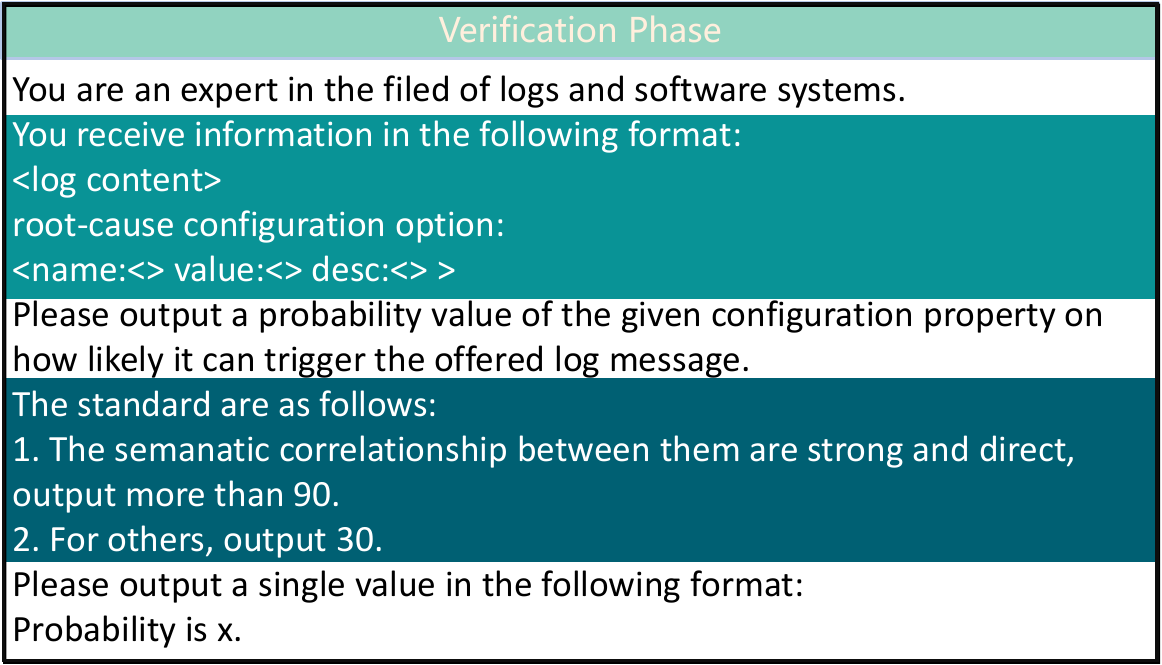}
    \vspace{-5pt}
    \caption{System Prompt in the Verification Phase}
    \label{pr:veri}
\end{figure}

\begin{figure}
    \vspace{-5pt}
    \centering
    \includegraphics[width=0.4\textwidth]{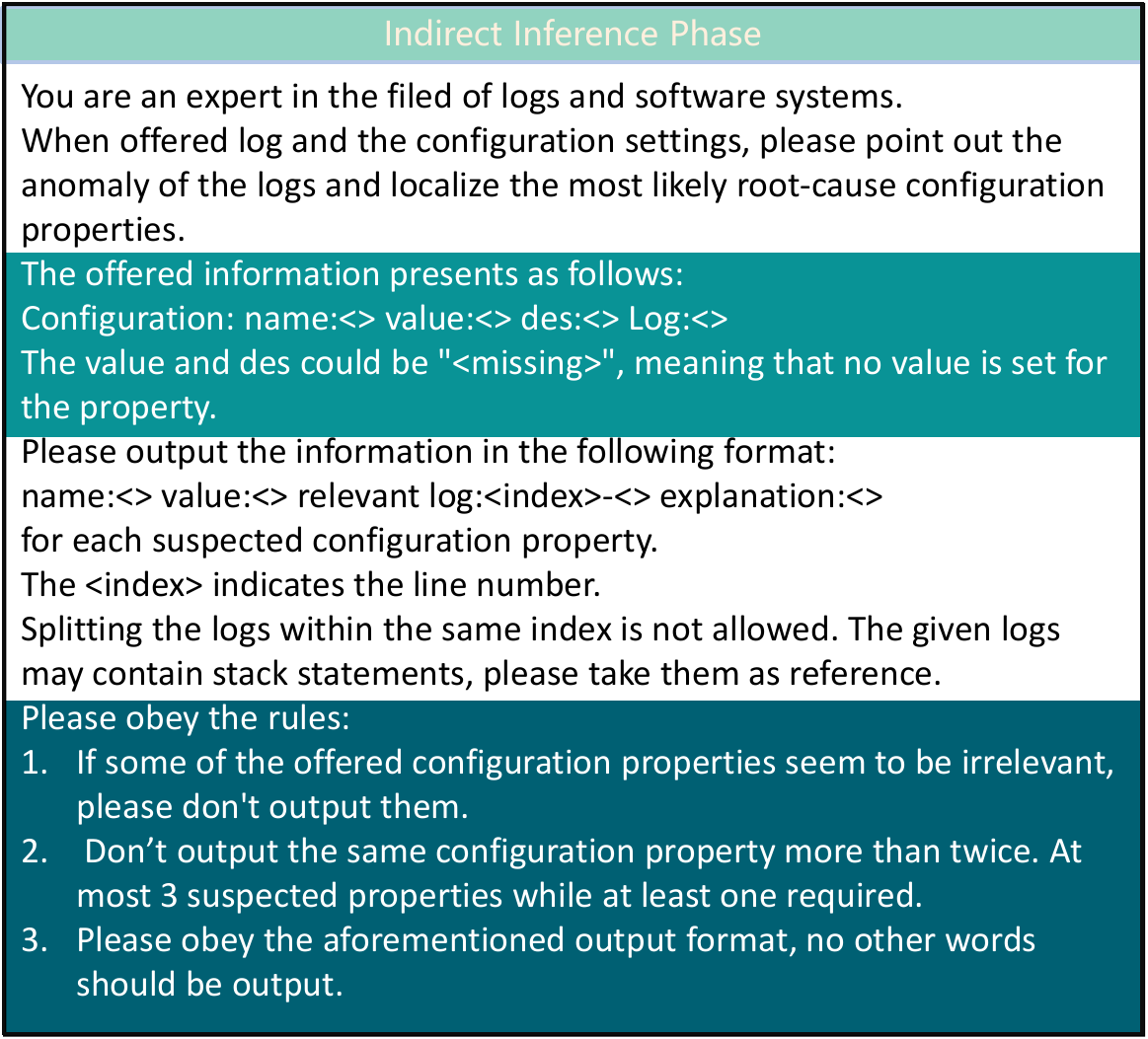}
    \vspace{-5pt}
    \caption{System Prompt in the Indirect Inference Phase}
    \vspace{-5pt}
    \label{pr:indirect}
\end{figure}

Following prior research~\cite{li2023exploring, le2023log, jiang2023LLMparser}, we choose GPT-4 Model~\cite{gpt4} (specifically, the fixed version, gpt-4-0613, to reproduce our results) as the default Large Language Model (LLM). GPT-4 Model~\cite{gpt4} is selected as the model for both the \textit{Verification Phase} and the \textit{Indirect Inference Phase}, following the precedent set by these previous studies.
To ensure the model consistently generates identical output for the same queries, thus guaranteeing reproducibility, we adjust the temperature setting to 0. The system prompts designed for the two phases are demonstrated in Figure~\ref{pr:veri} and Figure~\ref{pr:indirect}. For these processes, we utilize the public APIs provided by OpenAI~\cite{gpt4}.

\vspace{-5pt}
\subsection{Diagnosis Report Generation}
At the end of the localization procedure, we generate diagnosis reports for end-users. 
The report contains details about the configuration error triggers, including related log messages. 
We also demonstrate the explanations for the configuration error triggers to offer sufficient guidance to end-users. 
The explanation will show different messages if generated in different phases. Generated in the \textit{Verification Phase}, the explanation will show "value hits" or "name hits". While generated in the \textit{Indirect Inference Phase}, the explanation presents the illustration of GPT-4 Model~\cite{gpt4} on the suspected configuration error triggers. 
\section{Evaluation}\label{sec:evaluation}
To demonstrate the performance of our methodology, we follow four research questions to carry out our experiments. In particular, all the experiments are done on one single server equipped with Intel (R) Xeon (R) Gold 5218R CPU (2.1GHz) under the Ubuntu 22.04 LTS environment with 440GB physical memory.
\vspace{-5pt}
\subsubsection*{\textbf{RQ1:}} How accurate is the proposed methodology? 
\vspace{-5pt}
\subsubsection*{\textbf{RQ2:}} How effective of LogConfigLocalizer compared with other techniques?
\vspace{-5pt}
\subsubsection*{\textbf{RQ3:}} How effective of the Verification Phase?  
\vspace{-5pt}
\subsubsection*{\textbf{RQ4:}} How effective of the two parts of LLM interactions?

\subsection{Experiment Setup}
\subsubsection{Subject Software Systems}
We select Hadoop~\cite{hadoop}, a famous distributed big data framework, as the subject software system for several reasons: (1) Maturity: It has a matured configuration mechanism with a history spanning over fifteen years from version 0.10.1 to 3.3.6,~\cite{dong2016orplocator}; (2) Complexity: Its evolution into a complex ecosystem with over 100 related systems and numerous configuration properties makes its configuration space both large and intricate~\cite{dong2016orplocator,rabkin2011static,xu2023real,liu2023crsextractor}; 
(3) Popularity: It is widely used in academia and industry~\cite{rabkin2011static, liu2023crsextractor}. Therefore, Hadoop~\cite{hadoop} is representative for evaluating our methodology. 
\subsubsection{Benchmark Establishment}
As there is no existing log benchmark containing configuration errors for Hadoop~\cite{hadoop}, we establish such a benchmark using fuzzing technology on top of JQF~\cite{10.1145/3293882.3339002}, a coverage-guided fuzzer for Java programs. We also introduce mutations to configurations to simulate various real-time scenarios.

\subsubsection*{\textbf{(a) Sampling}} 
Due to the abundance of configuration properties in Hadoop~\cite{hadoop}, we randomly sample numeric configuration properties according to Finding 5. 
Particularly, the configuration properties are selected from the default configuration files. 
The overall number of the sampled configuration properties is 685 out of 1452.

\subsubsection*{\textbf{(b) Mutation Strategies}}
We develop two types of strategies for the value-level mutation: one adheres to the datatype specification, while the other deliberately violates it, as illustrated in Table~\ref{tab:mutate-value}. 
Unlike previous research centered on Configuration Error Injection Testing (i.e., CEIT)~\cite{li2024ecfuzz, 4630084, 10.1145/3460319.3464799}, we don't intend to inject specific types of configuration errors but to imitate the behaviors of end-users, for example, accidentally turn a positive value into a negative one without considering specific constraints of configuration properties.

Regarding the property-level mutation strategy, we randomly select one configuration property in the configuration space to replace the previous one in the former execution. 
This requires no effort to localize configuration error triggers in the \textit{Anomaly Inference Stage} since the configuration settings are accessed in both the direct and indirect inference phases. Therefore, we maintain the property-level mutation strategy unchanged but fabricate an additional configuration file for access. This involves injecting nine other configuration properties from the sampled configuration space with mutated values based on the same value mutation strategies.

\begin{table}[htb!]
  \caption{Strategies of Value Mutation}
  \vspace{-10pt}
  \label{tab:mutate-value}
    \begin{tabular}{c|c|c|c} 
    \toprule
    Data Type & Mutate Type  & Value Type & Range \\ 
    \hline
    \texttt{\multirow{5}{*}{Numeric}} & \multirow{3}{*}{Compliance} & Positive & $(0,MAX\_FLAOT)$   \\ 
    \cline{3-4}
    &  & Negative & $(MIN\_FLOAT,0)$   \\
    \cline{3-4}
    &  & Zero & $\{0\}$ \\
    \cline{2-4}
    & \multirow{2}{*}{Violation} & String & charset with 5 letters\\
    \cline{3-4}
    & & Empty & $\emptyset$\\
    \bottomrule
  \end{tabular}
\end{table}

\subsubsection*{\textbf{(c) Test Cases}} 
We utilize five workloads from HiBench~\cite{5452747}, a benchmark suite for big data framework, as test cases for log generation to simulate more realistic scenarios where the end-users run their application code within the software systems.
For each workload, we manually implement the test driver programs to activate the execution of fuzzing.
\subsubsection*{\textbf{(d) Execution}}
There are two modes to activate the fuzzing loop.
The default mode utilizes the conventional configuration settings to generate fault-free logs and the mutation mode executes with the mutated configuration settings (i.e., randomly select one configuration property with mutated value) to generate may-fault logs. 
The default mode runs for one hour, which is enough to generate multiple fault-free log files. We merge these fault-free log files into an integrated log file, parse it, and store the parsed log templates in the database. The mutated mode runs for eight hours to generate log files with a higher occurrence of configuration errors.
The details of the benchmark are presented in Table~\ref{tab:wk_info}. 
Concretely, we manually inspect each generated log file to determine if it is related to configuration errors. The statistics in the "C-Anomaly" column indicate the count of log files identified as related to configuration errors, while "w/o C-Anomaly" denotes those identified as configuration error-free. 

\begin{table*}
\small
  \caption{Benchmark. A single job executing the \texttt{pagerank} workload and \texttt{kmeans} workload produces two and four log files, respectively. We consider the log files generated in a job as a unified whole log file. The numbers in the brackets indicate the counts of these unified whole log files.  }
\vspace{-10pt}
  \label{tab:wk_info}
    \begin{tabular}{ccccccc} 
    \toprule
    {} & Mode & Gen-Log Files  & FuzzDuration & LogTemplate & C-Anomaly &  w/o C-Anomaly\\ 
    \hline
    \texttt{\multirow{2}{*}{wordcount}} & default & 132 & 1.000 & 128 & 0 &  132 \\ 
    & mutated & 1028 & 8.000 & 233 & 65  & 963 \\
    \cline{1-7}
    \texttt{\multirow{2}{*}{sort}} & default & 109 & 1.001 & 137& 0  & 109  \\ 
    & mutated & 459 &  8.000 & 201 & 151  & 308\\
    \cline{1-7}
    \texttt{\multirow{2}{*}{terasort}} & default & 129 & 1.020 & 128 & 0  & 129\\ 
    & mutated & 730 & 8.006 &215 & 227  & 503 \\
    \cline{1-7}
    \texttt{\multirow{2}{*}{pagerank}} & default & 92(46) & 1.582 & 130 & 0  &92(46)\\ 
    & mutated & 202(121) & 8.198 & 171 & 41  & 80  \\
    \cline{1-7}
    \texttt{\multirow{2}{*}{kmeans}} & default & 162(27) & 1.690 & 133 & 0  &162(27)\\ 
    & mutated & 226(48) & 8.186 & 157 & 14 &34 \\
    \bottomrule
  \end{tabular} \\ 
\end{table*}

\subsection{RQ1: How accurate is the proposed methodology?} 
To explore the accuracy of the proposed strategy, we apply LogConfigLocalizer on the established benchmark. Accuracy is calculated by the following formula for each test case: 
$$accuracy = \frac{counts\_{of}\_{correctly\_identified\_{test\_{cases}}}}{counts\_{of}\_{test}\_{cases}\_{flow\_into\_the\_ phase(stage)}}$$
Table~\ref{tab:accurate_info} demonstrates the average accuracy statistics. 

\begin{table}[]
  \caption{Accuracy. x-A denotes the accuracy of the x phase, for example, S1-A indicates the accuracy of the first stage and S2-D-A denotes the accuracy of the direct inference phase in the second stage.}
    \vspace{-10pt}
  \label{tab:accurate_info}
    \begin{tabular}{c|c|c|c|c} 
    \toprule 
    {} & S1-A & S2-D-A  & S2-I-A & S2-A  \\ 
    \hline
    \texttt{{wordcount}}& $100\% $ & $92.31\% $ &  $100\% $ & $100\% $  \\ 

    \texttt{{sort}} & $100\%$ & $100\%$ & $100\%$ & $100\%$  \\ 

    \texttt{{terasort}} & $100\%$ & $99.56\%$ & $100\%$ & $99.56\%$  \\ 

    \texttt{{pagerank}} & $100\%$ & $100\%$ & $100\%$ & $100\%$  \\ 

    \texttt{{kmeans}} & $100\%$ & $100\%$ & $88.89\%$  & $100\%$ \\ 
    \bottomrule 
  \end{tabular} \\
\end{table}






The proposed strategy achieves an average accuracy of 100\% in the \textit{Anomaly Identification Stage} across the five workloads, indicating the remarkable performance of \textit{Anomaly Identification Stage} to identify configuration-error-related cases.
Additionally, the \textit{Anomaly Inference Stage} attains an average accuracy of $99.91\%$. Both the direct and indirect inference phases exhibit high accuracy, with $98.37\%$ and $97.78\%$, respectively. This high accuracy underscores the reliability of the proposed strategy in localizing configuration errors.

In summary, LogConfigLocalizer excels in accurately pinpointing configuration error triggers in both the \textit{Anomaly Identification Stage} and the \textit{Anomaly Inference Stage}. The sole unsuccessful case in \texttt{terasort} workload results from a failure in the \textit{Verification Phase}, thus failing to proceed to the \textit{Indirect Inference Phase}. Nonetheless, when tested in the \textit{Indirect Inference Phase}, it effectively localizes the configuration error trigger. 
\vspace{-5pt}
\begin{tcolorbox}[frame empty]
\textbf{Answer to RQ1:} The proposed methodology attains a mean accuracy of up to $99.91\%$, affirming its feasibility and efficacy as a practical strategy for end-users to address configuration errors. 
\label{ans1}
\end{tcolorbox} 

\subsection{RQ2: How effective of LogConfigLocalizer compared with other techniques?}
We compare LogConfigLocalizer with ConfDiagDetector~\cite{zhang2015proactive}, which identifies a provided log message as diagnostic during a text analysis. The text analysis involves identifying the direct symptom within the message and gauging the semantic similarity between the message and the descriptions of the root-cause configuration properties, utilizing Natural Language Processing (NLP) techniques. 
It was not originally designed for localizing the configuration error triggers, however, its text analysis can be viewed as an approach to localizing the configuration error triggers. 

We utilize Hit Count as the metric to compare the performance of the tools. There are four Hit Counts: Name-Hit Counts, Value-Hit Counts, NLP-Hit Counts, and LLM-Hit Counts. The metrics represent the count of successfully identified log files.
The Name-Hit Count metric refers to detecting errors based on property names and the Value-Hit Count metric is based on property values.
The NLP-Hit Counts are employed for ConfDiagDetector, and the LLM-Hit Counts are used for LogConfigLocalizer. For instance, the LLM-Hit Counts indicate the count when the configuration error triggers are successfully localized in the \textit{Indirect Inference Phase}. The counting numbers are employed to assess the performance difference between LogConfigLocalizer and ConfDiagDetector.

\begin{figure*}[htbp]
    \centering
    \begin{minipage}[b]{\textwidth}
        \centering
        \includegraphics[width=\textwidth]{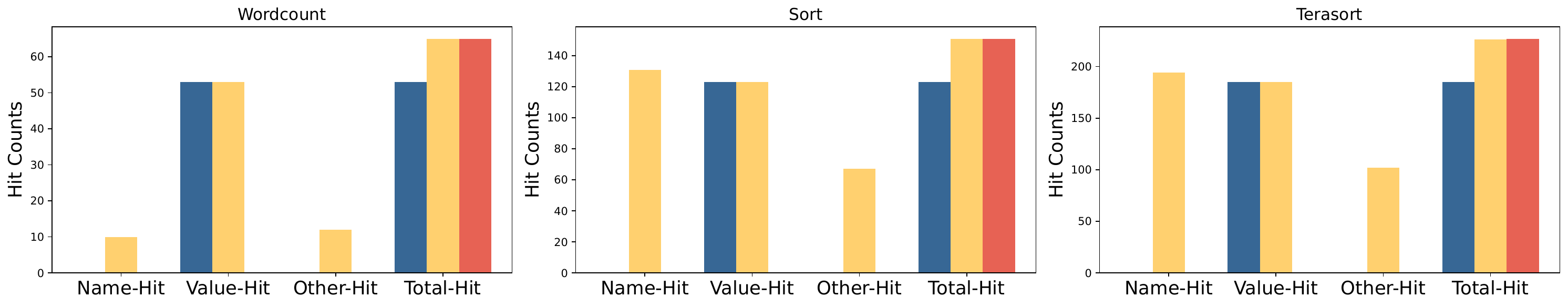}
        \label{fig:subfig1}
    \end{minipage}
    \hspace{-1pt}
    \begin{minipage}[b]{\textwidth}
        \centering
        \includegraphics[width=0.65\textwidth]{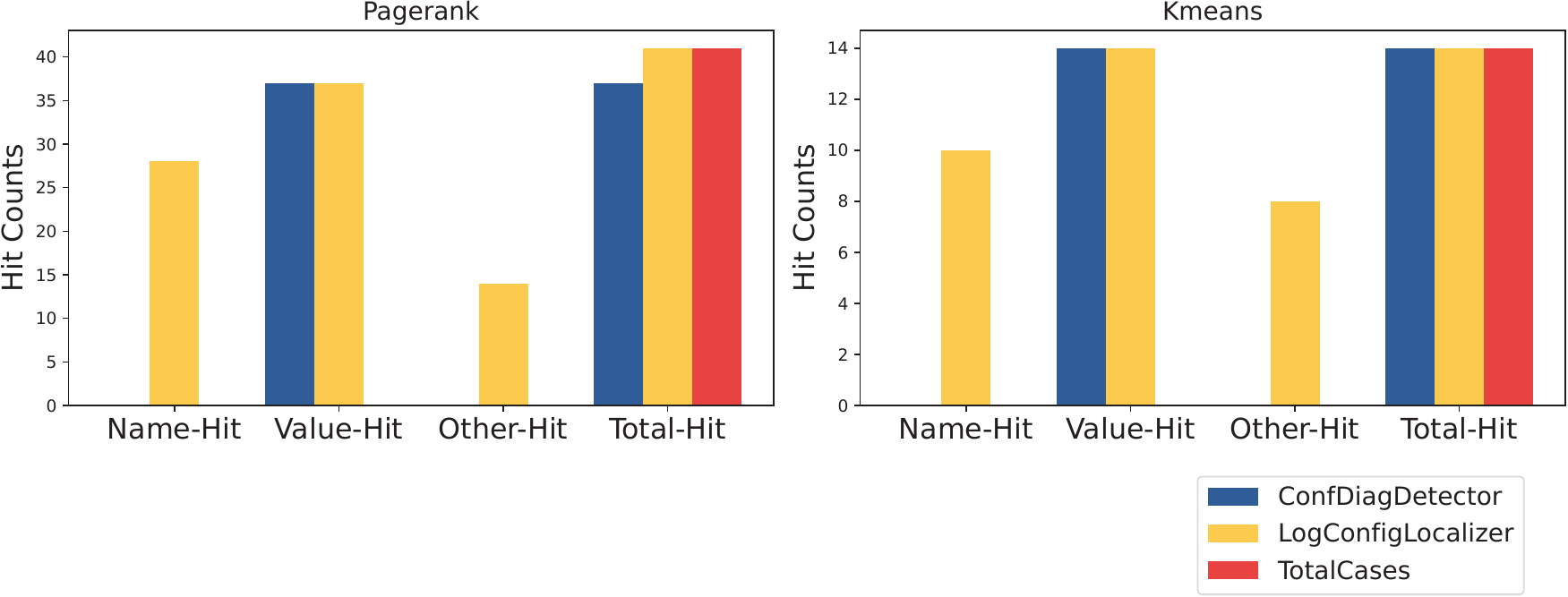}
        \label{fig:subfig2}
    \end{minipage}
        \caption{Comparison with ConfDiagDetector based on five workloads, the metric, Other-Hit, denotes LLM-Hit Counts for LogConfigLocalizer and NLP-Hit Counts for ConfDiagDetector respectively.}
    \label{fig:baseline}
\end{figure*}

We introduce the TotalCases Bar (Red-colored bars) in Figure~\ref{fig:baseline}, representing the total number of test cases in a workload, to demonstrate the accurate localization capabilities of the two tools.  
To assess the efficacy in localizing configuration errors, one can readily discern from Figure~\ref{fig:baseline} by comparing the Total-Hit Bar with the TotalCases Bar.  
LogConfigLocalizer nearly achieved hits for every test case across the five workloads, except for one in the \texttt{terasort} workload. 
In contrast, ConfDiagDetector's performance is inferior to LogConfigLocalizer, with accuracy on the five workloads below that of LogConfigLocalizer, especially with a significantly poorer performance in three of them.
For each test case in every workload, both LogConfigLocalizer and ConfDiagDetector successfully identify property values as their Value-Hit Counts remain constant across all workloads.
However, LogConfigLocalizer can capture both property names and values, whereas ConfDiagDetector consistently shows zero Name-Hit Counts across all workloads, as illustrated in Figure~\ref{fig:baseline}.
Additionally, ConfDiagDetector performs poorly when comparing the NLP-Hit Counts with the LLM-Hit Counts of LogConfigLocalizer. 
The ineffectiveness of the NLP technique in ConfDiagDetector may be attributed to the fact that even when log messages provide specific information indicating configuration error triggers, they often lack sufficient details regarding the corresponding descriptions. In contrast, LogConfigLocalizer excels at extracting key log messages and interpreting the underlying information within the logs.
\vspace{-5pt}
\begin{tcolorbox}[frame empty]
\textbf{Answer to RQ2:} LogConfigLocalizer outperforms ConfDiagLocalizer in terms of accuracy and exhibits a more versatile capability to localize configuration error triggers across different dimensions.
\label{ans2}
\end{tcolorbox} 


\vspace{-10pt}

\subsection{RQ3: How effective of the Verification Phase?}
To assess the effectiveness of the \textit{Verification Phase}, we introduce a variant, nv-LogConfigLocalizer (nv is short for no-verification), for comparison with the original version, o-LogConfigLocalizer.
The variant removes the \textit{Verification Phase} and considers configuration error triggers inferred in the \textit{Direct Inference Phase} as correct. In other words, the localization procedure enters the \textit{Indirect Inference Phase} only when the \textit{Direct Inference Phase} fails to identify any suspected configuration properties.
We utilize the accuracy metric to demonstrate the effectiveness of the \textit{Verification Phase} in ensuring the overall accuracy of the \textit{Anomaly Inference Stage}. The measurement is the same as that adopted in RQ1.
\begin{table}[htb!]
\small
  \caption{Comparison with nv-LogConfigLocalizer. The symbol "/" denotes no test cases flow into the specific phase.} 
  \label{tab:nv_variant_info}
    \begin{tabular}{c|c|c|c|c|c} 
    \toprule 
    {} &  S2-D-A & nv-S2-I-A & o-S2-I-A &  nv-S2-A & o-S2-A \\ 
    \hline
    \texttt{{wordcount}}  & $92.31\%$ & $/$  &  $100\%$  & $92.31\%$ & $100\%$ \\ 
    \texttt{{sort}}  & $100\%$ & $/$  & $100\%$  & $100\%$ & $100\%$   \\ 
    \texttt{{terasort}}  & $99.56\%$ & $100\%$  & $100\%$  & $99.56\%$ & $99.56\%$   \\ 
    \texttt{{pagerank}} & $100\%$ & $/$  & $100\%$  & $100\%$ & $100\%$    \\ 
    \texttt{{kmeans}} & $100\%$ & $/$  & $88.89\%$  & $100\%$ & $100\%$      \\ 
    \bottomrule
  \end{tabular} \\
\end{table}


Table~\ref{tab:nv_variant_info} demonstrates the results. The "nv-" column presents statistics from nv-LogConfigLocalizer, while the "o-" column represents the original version of LogConfigLocalizer. The unprefixed column shows identical statistics for both versions of LogConfigLocalizer.
In nv-LogConfigLocalizer and o-LogConfigLocalizer, the accuracy remains the same in the \textit{Direct Inference Phase}, as there are no changes in this phase. 
However, the absence of the \textit{Verification Phase} has an impact because, without it, there is no second opportunity to localize configuration error triggers if the \textit{Direct Inference Phase} fails. In the \texttt{terasort} workload, a test case proceeds directly to the \textit{Indirect Inference Phase}, resulting in successful localization by both o-LogConfigLocalizer and nv-LogConfigLocalizer. Yet, in five test cases from the \texttt{wordcount} workload, the verification failure prompts the localization procedure to enter the \textit{Indirect Inference Phase}, yielding higher accuracy ($100\%$ vs. $92.31\%$) in o-LogConfigLocalizer. Consequently, the lack of the \textit{Verification Phase} leads to an overall accuracy reduction.

\vspace{-5pt}
\begin{tcolorbox}[frame empty]
\textbf{Answer to RQ3:} The \textit{Verification Phase} is effective and essential for achieving higher accuracy in the \textit{Anomaly Inference Stage}. Without the \textit{Verification Phase}, the overall accuracy drops from $100\%$ to $92.31\%$ in the case of \texttt{wordcount} workload.
\label{ans3}
\end{tcolorbox}

\subsection{RQ4: How effective of the two parts of LLM interactions?} 
To comprehensively evaluate the effectiveness of LLM-interacted components, we introduce another variant, nl-LogConfigLocalizer (nl is short for no-LLM), to compare with the original version, o-LogConfigLocalizer.
The variant excludes the LLM interaction components from the \textit{Verification Phase} and the \textit{Indirect Inference Phase}. It utilizes a heuristic algorithm for the \textit{Verification Phase} and eliminates \textit{Indirect Inference Phase}. The heuristic algorithm posits that the configuration property matching with most anomalous log messages is the configuration error trigger. In this experiment, we take two metrics, the accuracy, and the false positive rate for our comparison. The false positive rate for each test case is calculated by the formula: $$FP = \frac{counts\_of\_suspected\_configuration\_properties-1}{counts\_of\_suspected\_configuration\_properties}$$
1 in the formula indicates the count of the ground truth.
Table~\ref{tab:nl_variant_info} displays the outcomes. 

A comparison between the third and fourth columns indicates that o-LogConfigLocalizer excels in minimizing the false positive rate in the \textit{Direct Inference Phase}, exhibiting a lower average false positive rate across all five workloads. Additionally, the incorporation of the \textit{LLM-based Indirect Inference Phase} in o-LogConfigLocalizer contributes to high accuracy.
\begin{table}[htb!]
\small
  \caption{Comparison with nl- variant} 
\vspace{-10pt}
  \label{tab:nl_variant_info}
    \begin{tabular}{c|c|c|c|c|c} 
    \toprule 
    {} &  S2-D-FP & nl-S2-V-FP & o-S2-V-FP &  nl-S2-A & o-S2-A \\ 
    \hline
    \texttt{{wordcount}}  & $70.03\%$ & $53.39\%$  &  $3.50\%$    & $92.31\%$ & $100\%$ \\ 
    \texttt{{sort}}  & $72.52\%$ & $17.58\%$  & $2.94\%$  & $100\%$ & $100\%$   \\ 
    \texttt{{terasort}}  & $71.75\%$ & $17.23\%$  & $2.00\%$  & $99.56\%$ & $99.56\%$   \\ 
    \texttt{{pagerank}} & $75.03\%$ & $14.10\%$  & $0$  & $100\%$ & $100\%$    \\ 
    \texttt{{kmeans}} & $75.29\%$ & $14.67\%$  & $10.0\%$  & $100\%$ & $100\%$      \\ 
    \bottomrule
  \end{tabular}
    \vspace{-10pt}
\end{table}
\vspace{-5pt}
\begin{tcolorbox}[frame empty]
\textbf{Answer to RQ4:} The introduction of LLMs guarantees the optimal effectiveness to reduce false positives and maintain a high overall accuracy.
\label{ans4}
\end{tcolorbox} 
\section{Practical Case Study}
We additionally perform a practical case study to demonstrate the feasibility of our methodology by localizing the configuration triggers on cases identified in the preliminary study.

For test case selection, we manually select 33 studied cases in the preliminary study to construct the practical benchmark. We employ three criteria for consideration: user-defined configuration settings, run-time logs under the user-defined configuration settings, and confirmed or resolved configuration settings that trigger the error. For the first two criteria, we manually review the content of the report to see if there is relevant information. As for the third criterion, we search for reports marked with confirmed or resolved flags. The practical benchmark covers path errors (6/33), classpath errors (4/33), boolean errors (5/33), numeric errors (10/33), and string errors (8/33).
The practical case study achieves an accuracy of $93.94\%$ (31/33),
which demonstrates the feasibility of our methodology.

To further demonstrate the effectiveness of our methodology, we categorize the correctly identified cases into three types, expanding upon the two aforementioned cases (i.e., the \textit{direct flow} and the \textit{complete flow}) described in Section \ref{sec:two_cases}. 
The three types include the \textit{fast flow}, \textit{direct flow}, and \textit{complete flow}. The \textit{fast flow} indicates a case successfully proceeding through the \textit{Direct Inference Phase}, passing the \textit{Verification Phase}, and skipping the \textit{Indirect Inference Phase}. 
Approximately $71\%$ (22/31) of the cases belong to either the \textit{fast flow} or the \textit{complete flow} type. This indicates that the \textit{Direct Inference Phase} effectively performs its intended function to pinpoint the suspected configuration error triggers, in the majority of cases.
However, about $73\%$ (16/22) of them fail to pass the \textit{Verification Phase}, consequently diverting the localization procedure towards the \textit{Indirect Inference Phase}.
This indicates that additional noise is introduced in real-world scenarios, highlighting the significance and necessity of introducing the \textit{Verification Phase} to ensure the overall accuracy of the \textit{Anomaly Inference Stage}. Moreover, the false positive ratio decreased from $0.43$ to $0.08$ due to the inclusion of the \textit{Verification Phase}.
In addition, about $27\%$ of the cases belong to the \textit{direct flow} type. 
This indicates the limitation of the \textit{Direct Inference Phase} in more realistic scenarios. 
However, we can leverage the \textit{Indirect Inference Phase} to localize the configuration error triggers.

\begin{figure}
    \centering
    \includegraphics[width=0.45\textwidth]{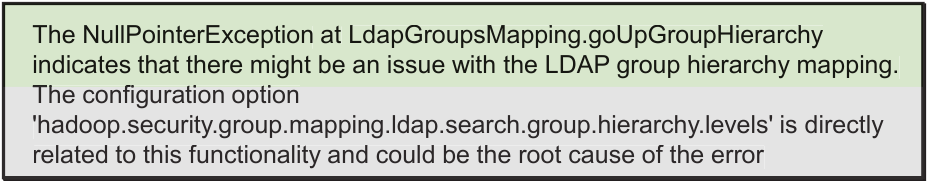}
    \vspace{-10pt}
    \caption{GPT-4 Model Explanation for Configuration Error Trigger}
    \label{prac_explanation}
    \vspace{-5pt}
\end{figure}

In addition, we take the first case and the last case shown in Figure~\ref{fig:two_tyoe_log_sym} to exemplify the proposed strategy.
The first case shown in the orange box\footnote{Original report: \url{https://issues.apache.org/jira/browse/HADOOP-134}.} belongs to the \textit{fast flow} type. During the \textit{Direct Inference Phase}, six entries are generated. The culprit, \texttt{mapred.local.dir}, is one of them due to the property name matching strategy, while the other five false positives are identified through value matching.
For example, the fabricated property, \texttt{dfs.datanode.du.reserved.pct} mutated with a value 0, matches the "[kry1040/72.30.116.100:50020]" string inside the log messages.
When the accordingly entries of the false positives are passed to the GPT-4 Model~\cite{gpt4}, it outputs 30 for each of them while offering a relatively high score, 95, to the entry of \texttt{mapred.local.dir}.
The case in the blue box\footnote{Original report: \url{https://issues.apache.org/jira/browse/HADOOP-18821}.} falls into the \textit{direct flow}, indicating no entry is generated during the \textit{Direct Inference Phase}. In the \textit{Direct Inference Phase}, we minimize false positives by excluding the analysis of stack statements. Consequently, in this case, the \textit{Direct Inference Phase} only receives the content "Error java.lang.NullPointerException," without any numeric values or other informative items. In the \textit{Indirect Inference Phase}, the GPT-4 Model~\cite{gpt4} identifies three potential configuration error triggers. It accurately identifies the most likely one, and the corresponding explanation is provided in Figure~\ref{prac_explanation}.

\section{Related Work}\label{sec:related} 
\subsection{Configuration Error Diagnosis}
Configuration errors are notorious and troublesome anomalies, prompting the proposal of numerous tools and methodologies to address them~\cite{6899207,7467302,zhang2015proactive,xu2016early,rabkin2011static,zhang2013automated,yuan2010sherlog,wang2018misconfdoctor,xu2023real,zhou2021confinlog}. In general, approaches to solving the problem can be categorized into two types: one that leverages machine learning techniques and another that exploits program analysis techniques.

Recent years have seen the rapid development of machine learning techniques. Xia et al.~\cite{6899207} utilized various text mining techniques to dig into the bug reports to tell whether a given bug report contains configuration errors or not. Similarly, Xu et al. proposed EFSPredictor~\cite{7467302}, a tool equipped with various feature selection approaches, to build a prediction model for the same target. Zhang et al.~\cite{zhang2015proactive} adopted the NLP technique to diagnose whether the output of the software systems implied the configuration error triggers. 
It's also a feasible way to explore deep into the source code. Ariel et al.~\cite{rabkin2011static} used static analysis to extract the configuration options for configuration debugging. Xu et al. introduced PCHECK~\cite{xu2016early} to automatically analyze the source code and generate configuration checking code to prevent configuration errors in advance. Zhang et al. presented ConfDiagnoser~\cite{zhang2013automated} using static analysis, dynamic profiling, and statistical analysis to localize the configuration error triggers. 

Beyond the aforementioned work, there are also associated research endeavors leveraging logs as auxiliary tools for diagnosing configuration errors. Similar to ConfigDiagDetector~\cite{zhang2015proactive}, Zhou et al.~\cite{zhou2021confinlog} utilized NLP technique to capture the NLP Patterns from official documents of software systems with log messages captured from source code and applied a pattern matching task to capture configuration constraints. 
Yuan et al. proposed SherLog ~\cite{yuan2010sherlog} to diagnose a production run failure by utilizing run-time logs and source code to infer the control and data flow of a failed execution.
Xu et al. put forward a real-time configuration error diagnosis method for software of AI server infrastructure~\cite{xu2023real}. The proposed framework requires source code to construct Abstract Syntax Trees and System Dependency Graphs to match the structured log templates extracted from real-time logs. 
Besides, Wang et al. presented MisconfDocotor~\cite{wang2018misconfdoctor}, a tool utilizing exception log features extracted from misconfiguration injection to identify configuration errors. 
The cited studies necessitate both source code and run-time logs to diagnose configuration errors, with a focus on run-time logs for control flow information, neglecting semantic content. 
Conversely, LogConfigLocalizer requires no access to source code, identifying abnormal logs and pinpointing configuration errors solely through semantic information extraction.
Moreover, instead of solely focusing on exception logs as MisconfDocotor~\cite{wang2018misconfdoctor} does, LogConfigLocalizer takes advantage of anomaly degree calculation for informative logs selection.

The proposed LLM-based two-stage strategy neither requires the source code of the software systems nor calls for data mining techniques or NLP techniques to dig into the logs. Meanwhile, it requires no misconfiguration error injection efforts. Therefore, it's more resource-saving and time-saving for end-users to adopt when encountering configuration errors. 
\vspace{-10pt}
\subsection{Log Analysis} 
Logs are valuable and informative outputs of software systems, hence, there is an abundance of research work targeted log analysis~\cite{huo2023autolog,huo2022logvm,du2017deeplog,zhu2023loghub,zhang2019robust,8903291,huo2023evlog}. 

To begin with, logging statement generation is crucial for the downstream log analysis task. 
Li et al. presented SCLogger~\cite{li2024go} to generate contextualized logging statements with static contexts. Yuan et al. put forward LogEnhancer~\cite{yuan2012improving} to enhance log messages based on source code for failure diagnosis.
Prior to log analysis, it is often necessary to utilize log parsing technologies to preprocess the raw log messages~\cite{zhang2019robust}. Huo et al. proposed Semparser~\cite{huo2023semparser}, a semantic-based parser to extract both explicit and implicit semantics of logs. Yu et al. introduced Log3T~\cite{yu2023log} to support new log types inside new-coming logs based on a transformer encoder-based model.
Logs have been utilized for anomaly detection for long. Du et al. proposed DeepLog~\cite{du2017deeplog}, a deep neural network model exploiting Long Short-Term Memory (LSTM) to detect anomalies based on the recognized log patterns. Zhang et al. introduced LogRobust~\cite{zhang2019robust}, utilizing an attention-based Bi-LSTM model for unstable log events and sequences to detect anomalies. Yang et al. presented nLSAlog~\cite{8903291}, an anomaly detection framework taking log sequences as sources to detect anomalies in Intelligent Transportation Systems based on the LSTM model and the self-attention mechanism. Previous research highlights the reliance on anomaly detection in log analysis on deep learning models, which require high-quality pre-collected and labeled datasets. The effectiveness of these methods hinges on dataset quality, a time-consuming process. Thus, a user-friendly and time-efficient anomaly identification stage, integrating rule-based strategies to assist end-users in detecting configuration errors, offers a novel approach compared to traditional methods.

LLMs have demonstrated significant capabilities in various domains, including fuzzing, type inference, and more~\cite{xia2023universal,chen2023empowering,kang2023preliminary,peng2023generative,li2024go,xu2024divlog}. The integration of LLMs with log analysis is also a growing area of research. Li et al.~\cite{li2023exploring} explored the performance of LLMs on the automated logging statement generation practice. Xu et al.~\cite{xu2023unilog} proposed UniLog to automatically decide where and what to log based on the in-context learning paradigm utilized in LLMs. Le et al.~\cite{le2023log} investigated the power of ChatGPT in automated log parsing, and Jiang et al.~\cite{jiang2023LLMparser} introduced LLMParser, an LLM-based log parsing framework to achieve better performance on log parsing. However, many existing works primarily concentrate on automated logging practice and log parsing, neglecting the capability of LLMs to enhance log analysis for anomaly identification and inference.

In summary, the LLM-based two-stage strategy for localizing configuration errors through logs is user-friendly compared to existing works using log analysis for anomaly detection. Additionally, we make a significant advancement by employing the robust natural language understanding capabilities of LLMs to interpret logs, thereby facilitating the localization of configuration errors.
\section{Conclusion}\label{sec:conclusion} 
Configuration error remains a challenging problem for both experienced maintainers and novice end-users, especially in the scenario of inaccessible source code. Given that logs are an easily accessible resource for most end-users, we conduct a preliminary study and outline the challenges and opportunities to apply log analysis to localize configuration errors. We present an LLM-based two-stage strategy via log analysis based on the insights from the preliminary study. To our knowledge, this is the first work to localize the root-cause configuration properties for end-users based on LLMs and logs. We implement LogConfigLocalizer based on the design of the strategy and show its effectiveness, accuracy, and feasibility via evaluations and a practical case study. We believe our work can alleviate the burden of end-users and facilitate a more streamlined use of configurable software systems.
\section*{Acknowledgements} 
 We appreciate all the anonymous reviewers for their valuable and practical comments. We also extend our gratitude to Zhuhan Dai, Weifeng Hu, Chengpeng Hu, Jiahao Cui, and Biao Zhu for their invaluable early feedback on the draft of this work. The work described in this paper was supported by the National Natural Science Foundation of China (No. 62202511) and the Guangdong Basic and Applied Basic Research Foundation (2022A1515011713). 
\newpage

\balance
\bibliographystyle{ACM-Reference-Format}
\bibliography{output}
\end{document}